\documentclass[fleqn,usenatbib]{mnras}

\usepackage[T1]{fontenc}
\usepackage{ae,aecompl}

\usepackage{graphicx}	
\usepackage{amsmath}	
\usepackage{amssymb}	
\usepackage[normalem]{ulem}
\usepackage{cuted}
\usepackage{nccmath}
\usepackage{addlines}

\usepackage{newtxtext,newtxmath}  

\newcommand{\mpc}{\,h^{-1}\,\rm cMpc} 
\newcommand{\lya}{Ly-$\alpha$}
\def\HI{\hbox{H$\,\rm \scriptstyle I$}}

\title[Lyman-$\alpha$ absorption signatures of protoclusters]{Searching for the shadows of giants II: the effect of local ionisation on the Lyman-$\alpha$ absorption signatures of protoclusters at redshift $z\sim2.4$}

\author[J. S. A. Miller, J. S. Bolton and N. A. Hatch]{
Joel S. A. Miller,\thanks{E-mail: joel.miller@nottingham.ac.uk}
James S. Bolton,
Nina A. Hatch
\\
School of Physics and Astronomy, University of Nottingham, University Park, Nottingham NG7 2RD, UK\\\\
}

\pubyear{2021}

\begin{document}
\label{firstpage}
\pagerange{\pageref{firstpage}--\pageref{lastpage}}
\maketitle

\begin{abstract}
Local variations in the intergalactic medium (IGM) neutral hydrogen fraction will affect the \lya\ absorption signature of protoclusters identified in tomographic surveys.  Using the IllustrisTNG simulations, we investigate how the AGN proximity effect and hot, collisionally ionised gas arising from gravitational infall and black hole feedback changes the Ly-$\alpha$ absorption associated with $M_{z=0}\simeq10^{14}\,M_\odot$ protoclusters at $z\simeq2.4$.   We find that protocluster galaxy overdensities exhibit a weak anti-correlation with \lya\ transmission in IGM transmission maps, but local \HI\ ionisation enhancements due to hot $T>10^{6}\rm\,K$ gas or nearby AGN can disrupt this relationship within individual protoclusters.   On average, however, we find that strong reductions in the IGM neutral fraction are limited to within $\lesssim 5h^{-1}\,\textrm{cMpc}$ of the dark matter haloes.  Local ionisation enhancements will therefore have a minimal impact on the completeness of protocluster identification in tomographic surveys if smoothing \lya\ transmission maps over scales of $\sim4 h^{-1}\,\textrm{cMpc}$, as is typically done in observations.  However, if calibrating the relationship between the matter density and \lya\ transmission in tomographic maps using simple analytical models for the \lya\ forest opacity, the presence of hot gas around haloes can still result in systematically lower estimates of $M_{z=0}$ for the most massive protoclusters.
 
\end{abstract}



\section{Introduction}

A fundamental prediction of $\Lambda\rm CDM$ cosmogonies is that galaxy clusters are built from the assembly of lower mass haloes such as galaxy groups and isolated galaxies \citep[e.g.][]{WhiteFrenk1991}. At low redshifts, clusters are single dark matter haloes, filled with massive, evolved, early-type galaxies orbiting the brightest cluster galaxy. The progenitor of this halo at redshift $z > 2$ is a diffuse collection of smaller halos spread over tens of comoving Mpc, all of which are rapidly growing and merging \citep{Chiang2013AncientProto-clusters, Muldrew2015WhatProtoclusters}. These ensembles of gravitationally bound but not yet virialised structures are known as protoclusters \citep{Overzier2016TheProtoclusters}.  They are the highest density regions in the early Universe and therefore are the most active sites of structure assembly. 

Traditionally, protoclusters are located as galaxy overdensities in either photometric \citep{Daddi2009TwoRedshifts, Chiang2014DiscoveryCosmos} or spectroscopic \citep{Steidel2005SpectroscopicRedshift, Cucciati2014DiscoveryVUDS, Chiang2015SurveyingHETDEX, Lemaux2017TheZ4.57, Harikane2019SILVERRUSH.Galaxies} surveys. Hundreds of protocluster candidates have now been identified as regions of high galaxy density on scales of a few to tens of arcminutes \citep{Wylezalek2013GalaxySpitzer, Toshikawa2018GOLDRUSH.Area}. A small but growing number of protoclusters have been selected not by their galaxy properties, however, but by their gaseous properties from the dominant baryonic component that lies in the intergalactic/intra-protocluster medium. X-ray and mm-wavelength observations -- the latter via the Sunyaev-Zel’dovich effect \citep{Sunyaev1972} -- are used to identify the $T\sim10^7$\,K intracluster medium within massive collapsed clusters and groups \citep{Ebeling_2010,Finoguenov_2010, Bleem_2015}, leading to the detection of clusters up to redshifts of $z=1.7$ \citep{Strazzullo_2019}. Yet the majority of the gas by volume within protoclusters has not yet been shock heated to X-ray emitting temperatures and instead has $T\sim10^4$\,K \citep{Miller2019SearchingAbsorption}.

Fortunately, neutral hydrogen in the intergalactic medium (IGM) also traces the underlying dark matter structure closely on scales $\gtrsim 1h^{-1}\rm\,cMpc$ \citep{Croft2002,Viel2004InferringSpectra}. Protoclusters at redshift $z>2$ have sizes of $10-50h^{-1}\rm~cMpc$ \citep{Muldrew2015WhatProtoclusters} so their large overdensities may be traced by neutral hydrogen.  This neutral hydrogen can be detected in the spectra of background quasi-stellar objects (QSOs) as a series of Ly$\alpha$ absorption lines known as the Ly$\alpha$ forest \citep{Rauch1998TheObjects}.  Coherent, large scale decrements in the Ly$\alpha$ forest transmitted flux within individual QSO spectra may then correspond to intergalactic \HI\ associated with significant mass overdensites.  In particular, \citet{Cai2017MAppingZ=2.32} have used spectral regions with strong \lya\ transmission decrements over a scale of $15h^{-1}\rm\,cMpc$ -- which they name Coherently Strong Ly$\alpha$ Absorption systems (CoSLAs) --  to locate a protocluster at $z=2.3$ \citep[see also][]{Cai2016MAppingMethodology,Zheng_2021,Shi2021}.  However, using cosmological hydrodynamical simulations, \citet{Miller2019SearchingAbsorption} (hereafter Paper I) also showed that such CoSLAs are rare\footnote{On analysing mock \lya\ forest spectra drawn through the $(80h^{-1}\rm\,cMpc)^{3}$ Sherwood simulation volume \citep{Bolton2017The5} with an average transverse separation of $0.75h^{-1}\,\rm cMpc$, only $\sim 0.1$ per cent of sight-lines exhibited a CoSLA (Paper I).} and are not an exclusive probe of protoclusters.  It is possible to adopt a rather strict (and model dependent) CoSLA detection threshold that removes contamination from coherent structures originating in the diffuse IGM, but any such protocluster sample is then incomplete.

An alternative technique for protocluster identification that also makes use of intergalactic \lya\ absorption is IGM or Ly$\alpha$ forest tomography \citep{Pichon_2001,Caucci_2008,Lee2014Observational2,Stark2015ProtoclusterMaps,Horowitz_2019,Porqueres_2020,Li2021}.  If a sufficient number of individual \lya\ forest sight-lines sample a given volume, a three dimensional map of the \lya\ transmission from the IGM may be reconstructed.  A large sample of QSO sight-lines can be used for this purpose \citep[see e.g.][]{Ravoux_2020}, but recent observational advances have also allowed spectra from background star-forming galaxies to be used, thus providing the high density of sight-lines needed to reconstruct the \lya\ transmission on scales of a few comoving Mpc \citep{Lee2016ShadowField,Lee2018First2.55,Mukae20203DTechnique,Newman2020LATIS:Survey}.  This approach has been successfully used to locate dense structures in the early Universe -- some of which are expected to be protoclusters.  Furthermore, combining these new tomographic \lya\ transmission maps with coeval galaxy surveys provides a powerful insight into the galaxy-IGM connection at $z>2$, and can improve the accuracy of the tomographic reconstruction of the underlying density field \citep{Mukae2017CosmicField,Momose_2020,Mukae_2020LAE,Liang_2021,Horowitz_2021}.

A key component in all of these recent IGM tomography studies are numerical simulations of the \lya\ transmitted flux; these are used to translate the 3D reconstruction of the \lya\ transmitted flux into the underlying matter density.  The most common approach used to create simulated \lya\ tomographic maps is to apply the fluctuating \citet{GunnPeterson1965} approximation (FGPA) to the density field from large collisionless dark matter simulations \cite[]{Stark2015ProtoclusterMaps,Lee2016ShadowField,Newman2020LATIS:Survey}. The FGPA assumes that the baryons trace the dark matter density field (modulo a correction for smoothing on the Jeans scale), that the neutral hydrogen is in photo-ionisation equilibrium with a spatially uniform UV background, and there is a single, power-law relationship between the gas density and temperature \citep[see e.g.][]{Rauch1998TheObjects,Becker2015RV}.  These assumptions are usually excellent ones when modelling the diffuse IGM at low densities, $\Delta=\rho/\langle \rho \rangle \lesssim 10$

It is well known, however, that neutral hydrogen (and hence also the \lya\ transmission) is not a completely unbiased tracer of the underlying density field, particularly in highly overdense regions.  More specifically, the FGPA will no longer hold for: (i)  gas that is hot and predominantly collisionally ionised, either due to shocks from gravitational infall or energetic feedback from supernovae driven winds and/or black hole accretion, (ii)  high density gas that is self-shielded to Lyman continuum photons, and (iii) local enhancements in the otherwise spatially uniform metagalactic UV background due to the presence of bright, rare sources (i.e. the proximity effect).  Indeed,  the presence of hot, highly ionised gas was suggested by \citet{Lee2016ShadowField} as an explanation for the lack of a strong \lya\ transmission decrement associated with a galaxy overdensity in their tomographic maps at $z\simeq 2.3$.  \citet{Mukae20203DTechnique} also demonstrated a spatial offset of $\sim 3$--$5h^{-1}\rm\,cMpc$ between \lya\ emitting galaxies and the minimum \lya\ transmission in their tomographic reconstruction around the MAMMOTH-1 nebula.   These authors suggested that local fluctuations in the ionising background may explain this offset, by changing the distribution of neutral hydrogen in the surrounding IGM \citep[see also][for a similar result obtained from the cross-correlation of galaxies and \lya\ tomographic maps]{Momose_2020}.

In this paper we investigate this issue of ``ionisation bias'' further using state-of-the-art hydrodynamical simulations from the IllustrisTNG project.
We explore how local ionisation variations in the IGM --  either due to the presence of hot, collisionally ionised gas or the QSO proximity effect -- impact on the detectability of protoclusters with Ly$\alpha$ forest tomography.    Furthermore, we assess how these variations may affect the relationship between the \lya\ transmission and the distribution of coeval galaxies, and how the assumption of the FPGA may bias constraints on protocluster mass.   The goal of this work is not to test the efficacy of tomographic reconstruction techniques; this is already discussed in the literature in some detail \citep[see e.g.][]{Stark2015ProtoclusterMaps,Horowitz_2019,Porqueres_2020,Li2021}.   In this work, rather than using a full forward model, we instead create idealised, noiseless \lya\ transmission maps by degrading our simulations to match the resolution of the tomographically reconstructed observations from  \citet{Lee2018First2.55} and \citet{Newman2020LATIS:Survey}.  The advantage of this approach is that it allows us to isolate the effect of astrophysical systematics from any uncertainties associated with the reconstruction methodology.


In Section \ref{sec:Sims} we introduce the hydrodynamical simulations and local ionisation models used throughout this work, and then examine the expected \lya\ transmission profiles around dark matter haloes in Section~\ref{sec:haloes}.  We discuss \lya\ transmission maps of protoclusters and their relationship with coeval Ly$\alpha$ emitting galaxies in Section \ref{sec:FvsGal}, and assess the role that local ionisation variations may play in \lya\ tomography measurements. Finally, we conclude in Section \ref{sec:Conclusions}.  Throughout this paper, we refer to comoving distance units using the prefix ``c" and to proper distance units using the prefix ``p".


\section{Simulating \texorpdfstring{L\lowercase{y}-$\alpha$}{} absorption from protoclusters}
\label{sec:Sims}
\subsection{Cosmological hydrodynamical simulations}
\label{ssec:SimComp}
In this work we shall primarily use the publicly available TNG100-1 simulation from the IllustrisTNG collaboration \citep{Nelson2019_TNG}.  IllustrisTNG has been performed using the moving-mesh hydrodynamics code \textsc{arepo} \citep{Springel2010EMesh}, and is described in detail in a series of five introductory papers \citep{Pillepich2018FirstGalaxies,Springel2018FirstClustering,Naiman2018FirstEuropium,Marinacci2018FirstFields,Nelson2018FirstBimodality}.  We use three further IllustrisTNG models with different box sizes and mass resolutions (TNG100-2, TNG100-3 and TNG300-1) to assess the numerical convergence of our results (see Appendix~\ref{app:Box&Res} for further details).  

In addition to the IllustrisTNG models, we also use the earlier Illustris-1 simulation \citep{Vogelsberger2014IntroducingUniverse,Nelson2015TheRelease} to assess the effect of a different sub-grid physics model on our results.  The key differences between IllustrisTNG and Illustris-1 are summarised in table 2 of \citet{Nelson2019_TNG}.  These include changes to the stellar and AGN feedback implementations, and the addition of ideal magneto-hydrodynamics in IllustrisTNG \citep{Pakmor2011MagnetohydrodynamicsGrid}.  There are also small differences in the $\Lambda$-CDM cosmological parameters used in the two models.   Importantly, however, the TNG100-1 initial conditions have the same random seed as Illustris-1, so we are able to directly compare the large scale structure of intergalactic gas in these models.  

All five of the simulations used in this work are summarised in Table~\ref{tab:Sims}.  For each simulation we use the snapshots and halo catalogues at $z=2.44$ and $z=0$. 

\begin{table}
	\centering
	 \caption{Hydrodynamical simulations used in this work.  The columns list, from left to right: the simulation name, the box size in $h^{-1}\, \rm cMpc$, the total number of gas cells and dark matter particles, and the typical dark matter particle and gas cell masses.  The IllustrisTNG simulations assume a \citet{PlanckCollaboration2015PlanckParameters} consistent cosmology, with $\Omega_{\rm m}=0.3089,\ \Omega_\Lambda=0.6911,\ \Omega_{\rm b}=0.0486,\ \sigma_8=0.8159,\ n_{\rm s}=0.9667\ \textrm{and}\ h=0.6774$.   The cosmological parameters used in the Illustris-1 simulation instead take values consistent with WMAP-9 \citep{Hinshaw2013Nine-yearResults}, giving $\Omega_{\rm m}=0.2726,\ \Omega_\Lambda=0.7274,\ \Omega_{\rm b}=0.0456,\ \sigma_8=0.809,\ n_{\rm s}=0.963\ \textrm{and}\ h=0.704$.  }
    \begin{tabular}{c|c|c|c|c}
    	\hline
    	Name        & Box size & $N_\textrm{\rm gas+DM}$ & $M_\textrm{dm}$ & $M_\textrm{gas}$  \\
        		    &[$h^{-1}\ $cMpc]&  & [$M_\odot$] & [$M_\odot$]  \\
        \hline
        TNG100-1    & 75 & $2\times1820^3$ & $7.50\times10^6$  & $1.40\times10^6$  \\
        Illustris-1 & 75 & $2\times1820^3$ & $6.26\times10^6$ & $1.26\times10^6$  \\
		TNG100-2 & 75 & $2\times 910^3$ & $5.97\times10^7$ & $1.12\times10^7$\\
		TNG100-3 & 75 & $2\times 455^3$ & $4.78\times10^8$ & $8.92\times10^7$\\
		TNG300-1 & 205 & $2\times 2500^3$ & $5.90\times10^7$ & $1.10\times10^7$\\
		\hline
    \end{tabular}
    \label{tab:Sims}
\end{table}

\subsection{Protocluster identification}

Throughout this paper we take advantage of the ability of simulations to connect physical structures at different instances in time.  Following Paper I, we define protoclusters in the TNG100-1 model as the structures that form clusters with $M_{\rm z=0}\geq 10^{14}\, \rm M_\odot$ at redshift $z=0$. We identify the simulation resolution elements that belong to these protoclusters as being all those within friends-of-friends haloes with $M_{\rm z=0}\geq 10^{14}\, \rm M_\odot$ at redshift $z=0$. We then find these resolution elements at redshift $z=2.44$ and use their positions to compute the centre of mass of each protocluster and the radial extent around the centre of mass, $R_{95}$, that contains 95 per cent of the protocluster's $z=0$ mass.  This procedure yields a total of 22 protoclusters in the TNG100-1 volume at $z=2.44$. 

\begin{figure*}
	\includegraphics[width=\textwidth]{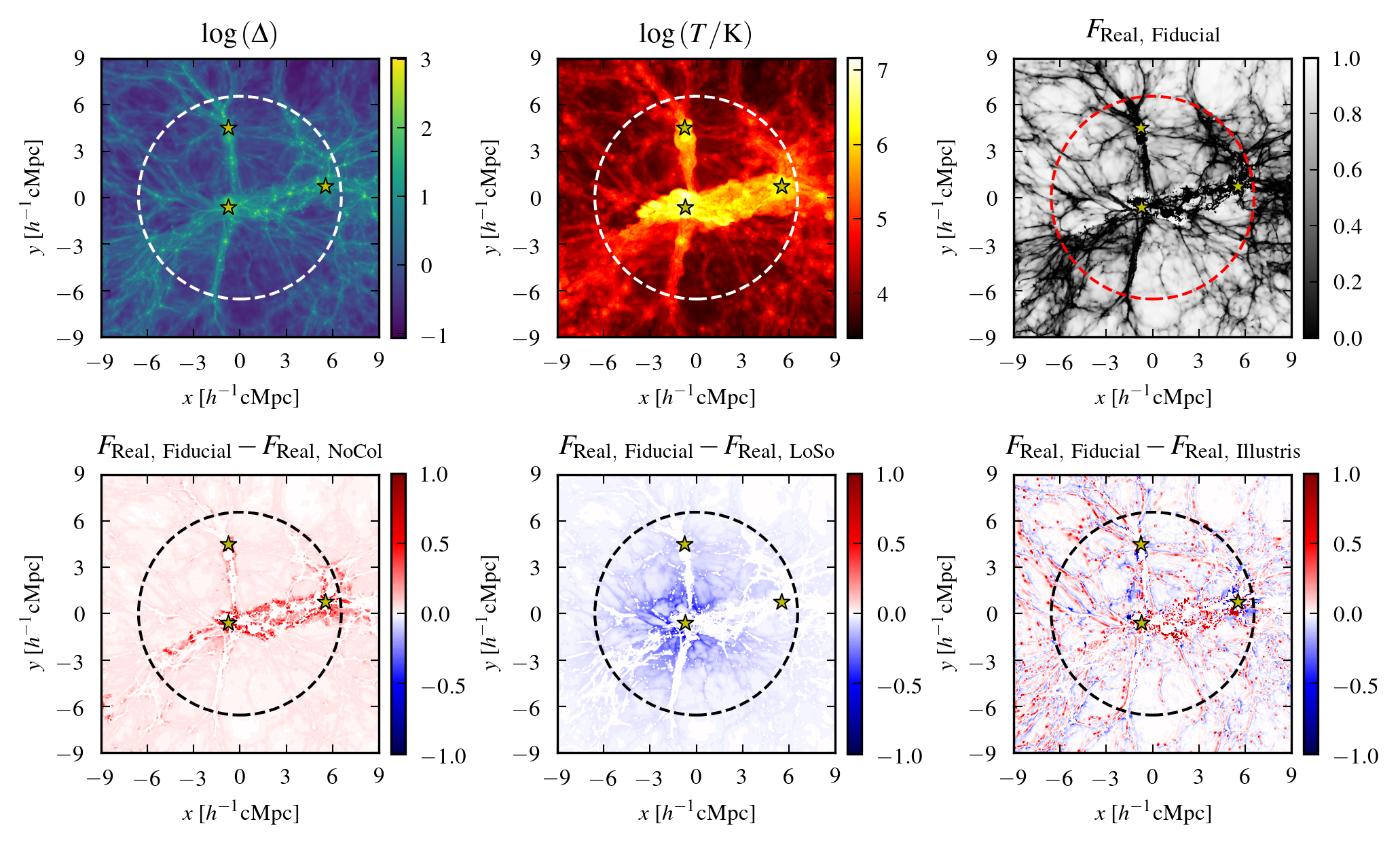}
	\vspace{-0.7cm}
    \caption{Top: A series of 2D projections of the gas overdensity (left), gas temperature (centre) and the real space \lya\ forest transmission $F_{\rm real}$ (right, see text for details) for a protocluster with $z=0$ mass $M_{\rm z=0}=10^{14.46}\,\rm M_{\odot}$ in the TNG100-1 simulation at $z=2.44$.   The slices are projected over a distance of $2~h^{-1}~\rm cMpc$ and are centred on the protocluster centre of mass. The dashed circle respresents $R_{95}$ for the protocluster, while the yellow stars denote the locations of haloes within the slice that are populated with AGN in our local source ionisation model (see Section~\ref{sec:ionmod} for details). Bottom: Slices showing the difference in $F_{\rm real}$ between our fiducial model and a model with no collisional ionisation or self-shielding (left), a model with local enhancements in the IGM ionisation state due to the proximity effect from AGN (centre), and the Illustris-1 simulation which uses a different sub-grid physics implementation compared to TNG100-1. Here red represents a larger \lya\ transmission (less absorption) in the fiducial model, while blue represents a smaller transmission (more absorption).}
    \label{fig:PC_Map}
\end{figure*}

An example of one such protocluster from the TNG100-1 simulation with $z=0$ mass $M_{\rm z=0}=10^{14.46}\, \rm M_{\odot}$ is displayed in the upper left and central panels of Fig.~\ref{fig:PC_Map}. A 2D projection of the logarithm of the normalised gas density, $\Delta= \rho/\langle \rho \rangle$, and logarithm of the gas temperature, $T$,  are shown within $\pm 1 \,h^{-1}\,\rm cMpc$ of the protocluster centre of mass.  The white dashed circle in each panel shows the radial extent of the protocluster, $R_{95}$.  As discussed in detail in Paper I, a wide range of protocluster morphologies are expected using our protocluster definition, where typically $R_{95}=5$--$10h^{-1}\rm\,cMpc$.  On average, the gas in protoclusters will exhibit slightly higher densities, temperatures and neutral hydrogen fractions compared to the surrounding IGM.

\subsection{Local ionisation models} \label{sec:ionmod}

The primary focus of this work is assessing the impact that local variations in the IGM ionisation state may have on the identification of protoclusters using \lya\ absorption.  We now turn to describing the three different IGM ionisation models we use for this purpose.

In our fiducial ionisation model we adopt a similar approach to Paper I and assume a spatially uniform UV background using the \citet{Faucher-Giguere2019AUpdate} synthesis model.  For reference, the \citet{Faucher-Giguere2019AUpdate} model has an \HI\ photo-ionisation rate $\Gamma_{\rm HI}=9.76\times10^{-13}\rm\,s^{-1}$ at $z=2.44$, which is consistent with independent constraints on $\Gamma_{\rm HI}$ from the \lya\ forest opacity \citep{BeckerBolton2013}.
We calculate neutral hydrogen fractions in each cell of the simulation under the assumption of ionisation equilibrium by using the coupled equations given by \citet{KWH96}, after updating the recombination and collisional ionisation rates to match those used by \citet{Bolton2017The5}.    We also use the \citet{Rahmati2013OnSimulations} prescription for self-shielding to obtain the correct incidence of absorbers that are optically thick to Lyman continuum photons (i.e. for $N_{\rm HI}\geq10^{17.2}\,\rm cm^{-2}$).  We have already verified in Paper I (see fig. 1 in that work) that this procedure reproduces the shape of the observed \HI\ column density distribution over the range $10^{12}\rm\,cm^{-2}\leq N_{\rm HI}\leq 10^{22}\rm\,cm^{-2} $ very well.

In addition to our fiducial model, we investigate two further, alternative ionisation models.  In the first we assume a spatially uniform UV background, but now ignore the effects of collisional ionisation and self-shielding on the neutral hydrogen fraction.  We achieve this by setting the collisional ionisation rates to zero and neglecting the \citet{Rahmati2013OnSimulations} correction when calculating the \HI\ fractions in each gas cell of the hydrodynamical simulations.    Collisional ionisation will be particularly important for the ionisation state of gas around haloes, where gas is heated to $T>10^{6}\rm\,K$ by gravitational infall and AGN or supernovae feedback (see e.g. the protocluster in the upper central panel of Fig.~\ref{fig:PC_Map}).  Neglecting collisional ionisation in these hot, dense regions will result in an overestimate of the \HI\ fraction, and hence an overestimate of the \lya\ optical depth associated with the gas.  By contrast, ignoring self-shielding will instead result in an underestimate of the number of rare, high column density absorption systems with $N_{\rm HI}\geq10^{17.2}\,\rm cm^{-2}$ that arise from cool, dense gas.  We refer to this model as ``No Collisional" -- shortened to \textit{NoCol} -- throughout this paper.   The \textit{NoCol} model is chosen to be similar (but not identical) to the fluctuating Gunn-Peterson approximation (FGPA) that has been commonly used in the recent literature to link the \lya\ optical depth to the underlying gas or dark matter density \citep[e.g.][]{Stark2015ProtoclusterMaps,Newman2020LATIS:Survey}.  The FGPA assumes photo-ionisation equilibrium in an IGM which follows a power-law temperature density relation, $T=T_{0}\Delta^{\gamma-1}$, which is a good approximation only for gas with $\Delta \leq 10$ at $z\simeq 2$ \citep[see e.g.][]{Rauch1998TheObjects}.  

Our second alternative ionisation model includes the effect of local enhancements in the IGM ionisation state due to quasars and active galactic nuclei \citep[i.e. the proximity effect,][]{Murdoch1986,Bajtlik1988}.  At the redshift we consider in this work, $z=2.44$, the mean free path of Lyman continuum photons is $\sim 300\rm\,pMpc$ \citep{Worseck2014}; on smaller scales the UV background is to a good approximation spatially uniform.  However, the presence of active galactic nuclei (AGN) in close proximity to protoclusters could mean the background photo-ionisation rate is significantly enhanced on scales up to a few proper Mpc in the vicinity of the AGN.

We model the effect of a local enhancement in the ionisation level of neutral hydrogen following the simple model described in \citet{BoltonViel2011}.  We populate the TNG100-1 simulation with AGN at $z=2.44$ by requiring the number of AGN in a comoving volume, $V$, satisfies
\begin{equation} N_{\rm AGN} = V \int_{L_{\rm min}}^{\infty} \phi(L_{\rm 1450})dL_{\rm 1450}, \end{equation}             
\noindent
where $\phi(L_{1450})$ is the AGN luminosity function from \citet{Kulkarni2019Evolution7.5} at $z=2.44$.  We assume a minimum luminosity of $L_{\rm min}=10^{43.2}\rm\,erg\,s^{-1}$, corresponding to an absolute AB magnitude $M_\textrm{1450}=-18$.    We assign a luminosity, $L_{1450}$, to each AGN by Monte Carlo sampling the luminosity function from \citet{Kulkarni2019Evolution7.5}, and then populate the simulation by assigning AGN to haloes in a one-to-one rank order fashion, such that the most luminous AGN resides in the most massive halo (i.e. we effectively assume an AGN duty cycle of one).  This yields $281$ AGN within the TNG100-1 volume, with a median $M_{1450}=-18.9$ and a minimum of $M_{1450}=-24.7$.

Next, for each AGN we assume the spectral energy distribution used by \citet{Kulkarni2019Evolution7.5},
\begin{equation} L(\nu) \propto \Bigg\{ \begin{aligned} &\, \nu^{-0.61} \, (912\rm \, \textup{\AA} < \lambda \leq 1450 \rm \, \textup{\AA}),\\
& \, \nu^{-1.70}\, (\lambda \leq 912 \rm \, \textup{\AA}). \end{aligned}
\end{equation}                   
\noindent
We then compute the specific intensity, $J({\bf r},\nu)$, of the ionising emission from the AGN on a $256^3$ grid, assuming each AGN emits isotropically and that the IGM is optically thin within the periodic simulation volume.  Hence
\begin{equation} J({\bf r},\nu)= \frac{1}{4\pi}\sum_{\rm i=1}^{\rm N_{\rm AGN}} \frac{L_{\rm i}({\bf r},\nu)}{4\pi|{\bf r}_{\rm i}-{\bf r}|^{2}}, \end{equation}
where $|{\bf r}_{\rm i}-{\bf r}|$ is the distance of the $i^{\rm th}$ AGN from ${\bf r}$.  Finally we compute the spatially varying photo-ionisation rate from the AGN by evaluating 
\begin{equation} \Gamma_{\rm HI}({\bf r})=\int_{\nu_{\rm HI}}^{4\nu_{\rm HI}}\frac{4\pi J({\bf r},\nu)}{h_{\rm P}\nu} \sigma_{\rm HI}(\nu)\,d\nu, \label{eq:PIrate} 
\end{equation}
\noindent
where $\sigma_{\rm HI}(\nu)$ is the photo-ionisation cross-section from \citet{Verner1996} and $\nu_{\rm HI}$ is the frequency at the hydrogen Lyman limit.  The photo-ionisation rate for each gas cell is then obtained by trilinear interpolation of the nearest $256^{3}$ grid points to the Voronoi cell centre.  If the photo-ionisation rate from  Eq.~(\ref{eq:PIrate}) exceeds the value from the \citet{Faucher-Giguere2019AUpdate} synthesis model at $z=2.44$ in any given cell, we use the former to calculate the ionisation fraction.  Throughout this work we shall refer to this as our ``local sources'' -- shortened to \textit{LoSo} -- model.  

In Fig. \ref{fig:PC_Map}, we perform an initial assessment of the effect of these ionisation models on the average \lya\ forest transmission.   We consider a region of width $\Delta R= 2\,h^{-1}\,\rm cMpc$ centred around the protocluster, and obtain an estimate of the real space transmitted flux, $F_{\rm real}$, from the column density, $N_{\rm HI}$, in each pixel following a similar approach to \citet{Kulkarni2015}, where
\begin{equation}
    F_{\rm real} = \exp\left(-\frac{3\lambda_{\rm Ly\alpha}^{3}\gamma_{\rm Ly\alpha}}{8\pi H(z)} \frac{N_{\rm HI}}{\Delta R} \right). 
    \label{eq:Freal}
\end{equation}
Here $\gamma_{\rm Ly\alpha}=6.265\times 10^{8}\rm\,s^{-1}$ is the \lya\ damping constant and $\lambda_{\rm Ly\alpha}=1216 \rm\, \textup{\AA}$.  This approximation ignores the effect of peculiar velocities and thermal broadening on the \lya\ opacity, and as a consequence it does not provide an accurate value for the average \lya\ transmission along a given line of sight.  However, it provides a convenient illustration of the relative effect of our ionisation models on the \lya\  opacity within protoclusters.

In the top right panel of Fig.~\ref{fig:PC_Map} we show F$_{\rm real}$ for the example TNG100-1 protocluster at $z=2.44$ in the fiducial ionisation model, whilst in the lower panels we show the difference in $F_\textrm{real}$ between the fiducial model and the \textit{NoCol} model (left), \textit{LoSo} model (centre), and the same region in the Illustris-1 simulation (right).  In the \textit{NoCol} model there is a decrease in the transmission from the  filaments within the protocluster, leading to relatively more transmission in the fiducial model. The filaments (which are most apparent in the upper panels of Fig.~\ref{fig:PC_Map}) span up to $\sim10\,h^{-1}\,\rm cMpc$ in length, with widths on the order of $\sim100\,h^{-1}\,\rm ckpc$, and  consist of overdense gas ($\Delta \sim 10$--$100$) at high temperatures ($T\sim10^{5}$--$10^{7}\, \rm K$).  This corresponds to gas that has been heated by shocks and outflows and is therefore collisionally ionised in the fiducial model.  Hence, we expect that ignoring hot, collisionally ionised gas will underestimate the \lya\ transmission from the gas in protoclusters.  By contrast, in the \textit{LoSo} model there is an increase in the \lya\ transmission in the protocluster  relative to the fiducial model,  with a magnitude that decreases radially and is generally more pronounced in cooler, less dense regions where photoionisation dominates.  Finally, comparing the different sub-grid physics implementation used in Illustris-1 to the fiducial TNG100-1 model (see Section~\ref{ssec:SimComp} for further details), we find the variation in  transmission along the filaments of the cosmic web is more complex.  Once again, these differences are driven primarily by changes in the thermal and ionisation state of the hydrogen gas.\footnote{This comparison is not exact, however, due to the slightly different cosmological parameters used in Illustris-1 and TNG-1 (see Table~\ref{tab:Sims})}  Note the transmission from the low density IGM with $\Delta \lesssim 1$ remains unchanged, however, as the gas in voids is largely unaffected by shocks, AGN or supernovae driven winds at $z=2.44$ \citep[e.g.][]{Theuns2002,Viel2013TheFlux}.

\subsection{Mock \texorpdfstring{L\lowercase{y}-$\alpha$}{} absorption spectra}
\label{sec:spec}

\begin{figure*}
	\includegraphics[width=\textwidth]{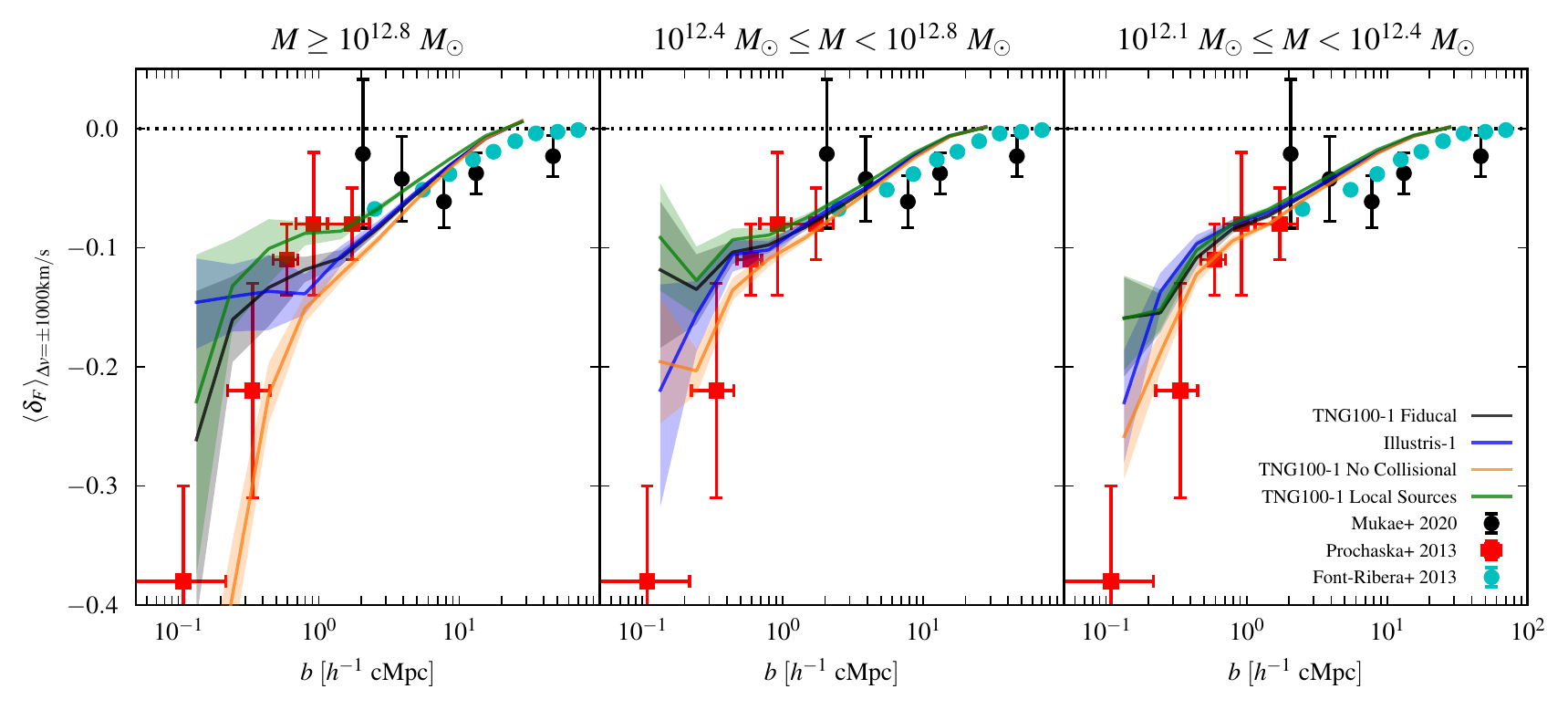}
	\vspace{-0.8cm}
    \caption{The median of the relative \lya\ transmission, $\delta_{\rm F}$, averaged over a velocity window of width $\Delta v =2000 \rm km\,s^{-1}$, as a function of impact parameter, $b$, around haloes in three mass bins at redshift $z=2.44$: $M\geq 10^{12.8}\,M_{\odot}$ (left), $10^{12.4}\,M_{\odot}\leq M < 10^{12.8}\,M_{\odot}$ (centre) and $10^{12.1}\,M_{\odot}\leq M < 10^{12.4}\,M_{\odot}$ (right).  A negative (positive) value of $\delta_{\rm F}$ corresponds to a decrease (increase) in the \lya\ transmission relative to the mean transmitted flux, $\langle F \rangle$, of the IGM.  The spatially uniform UV background model from the TNG100-1 simulation (black curves) is compared to the case ignoring collisional ionisation and self-shielding (orange curves, \textit{NoCol} model) and with the addition of a proximity effect due to AGN (green curves, \textit{LoSo} model). The corresponding haloes in the Illustris-1 simulation are shown by the blue curves.   The shaded regions bound 68 per cent of the distribution around the median, and are obtained by $10^{3}$ bootstrap samples of the transmission profiles around each halo.  For comparison, the data points and $1\sigma$ error bars display observational constraints on the \lya\ transmission around QSO host haloes from \citet{Font-Ribera2013TheBOSS} (cyan circles), \citet{Prochaska2013QuasarsQuasars} (red squares) and \citet{Mukae20203DTechnique} (black circles).}
    \label{fig:haloRadPhys}
\end{figure*}

In the remainder of this work we will analyse the \lya\ absorption associated with protoclusters using simulated \lya\ forest spectra.  We again follow the procedure described in Paper I, which we briefly repeat here.  Mock \lya\ absorption spectra are extracted from the simulations by assigning each Voronoi cell a smoothing length, $h_{\rm i}$, based on the cell volume, $V_{\rm i}$, such that
\begin{equation} h_{\rm i} = \left(\frac{3N_{\rm sph} V_{\rm i}}{4\pi}\right)^{1/3}. \end{equation}
We assume $N_{\rm sph}=64$ for the number of smoothing neighbours.  The interpolation scheme described by \citet{Theuns1998P3M-SPHForest} is then used to extract \lya\ optical depths using the Voigt profile approximation from \citet{Tepper-Garcia2006VoigtFunction}.   Unless otherwise stated, we also rescale the optical depths of each pixel in our mock spectra by a constant to match observational constraints on the \lya\ forest effective optical depth, $\tau_{\rm eff}=-\ln\langle F \rangle = 0.20$ at $z=2.4$, from \citet{Becker2013ASpectra}.  

The transmitted flux in each pixel is then given by $F = e^{-\tau}$, and we define the transmitted flux contrast, $\delta_{\rm F}$ as the relative transmission -- averaged over some velocity window of width $\Delta v$ -- around the IGM mean value
\begin{equation}
    \delta_{\rm F} = \frac{\langle F\rangle_{\Delta v}}{\langle F\rangle} - 1. 
    \label{eq:deltaf}
\end{equation}
\noindent
A negative (positive) value of $\delta_{\rm F}$ thus represents a decrease (increase) in the \lya\ transmission relative to the mean transmitted flux, $\langle F \rangle$, of the IGM.


\section{\texorpdfstring{L\lowercase{y}-$\alpha$}{} absorption profiles around haloes}
\label{sec:haloes}

We perform a consistency test of our mock \lya\ absorption spectra in Fig.~\ref{fig:haloRadPhys}, where we show the transmitted flux contrast for our different ionisation models around haloes in three mass bins:  $M\geq 10^{12.8}~M_\odot$ (left), $10^{12.4}~M_\odot \leq M<10^{12.8}~M_\odot$ (centre) and $10^{12.1}~M_\odot \leq M<10^{12.4}~M_\odot$ (right).  We select the mock spectra using a grid of sight-lines running the length of the simulation box in all three cardinal directions, with a mean transverse separation of $1.96\mpc$. We then calculate the mean transmission within a velocity window of 2000$\,{\rm km}\,{\rm s}^{-1}$, and bin the transmission in terms of the halo impact parameter, $b$.   The results are compared to observational measurements of $\delta_{\rm F}$ around QSOs from \citet{Mukae20203DTechnique}, \citet{Prochaska2013QuasarsQuasars} and \citet{Font-Ribera2013TheBOSS}. Note that we display the \citet{Mukae20203DTechnique} MAMMOTH1-QSO measurements only on scales above the resolution limit of their \lya\ tomographic maps.  The \citet{Font-Ribera2013TheBOSS} data correspond to the Baryon Oscillation Spectroscopic Survey (BOSS) QSO-Ly$\alpha$ cross-correlation measurement, and have been converted to $\delta_{\rm F}$ by \citet{Sorini2018ADistance}.

Several earlier studies have already discussed the level of agreement between hydrodynamical simulations and observations of the neutral hydrogen distribution around QSOs \citep[e.g][]{Fumagalli2014,Rahmati2015,FaucherGiguere2016,Meiksin2017GasSuite,Sorini2020Simba:Feedback,Nagamine2020ProbingJWST}. In general, differences in stellar and AGN feedback implementations, halo mass and numerical resolution all play an important role.  The differences we find here are consistent with earlier work, where the relative transmission at small scales, $b<0.5h^{-1}\rm\,cMpc$, in Illustris-1 (blue curves) and TNG100-1 (black curves) is larger than the observed relative transmission.  The relative difference between these two models, particularly in the $M\geq 10^{12.8}\rm\,M_{\odot}$ bin, is most likely associated with the more aggressive AGN feedback implementation within Illustris-1, which leads to more hot, collisionally ionised gas.  

Recently, however, \citet{Sorini2020Simba:Feedback} have found very good agreement between the \citet{Prochaska2013QuasarsQuasars} data and \textsc{SIMBA} simulations \citep{Dave2019} on small scales, suggesting that the choice of stellar feedback model plays a key role in reproducing the observations \citep[see also][]{FaucherGiguere2016}.   On larger scales ($b>1h^{-1}\rm\,cMpc$), the level of agreement we find with the \citet{Font-Ribera2013TheBOSS} data is similar to \citet{Sorini2020Simba:Feedback}, who suggest the overprediction of the relative transmission may be due to the small box size ($50h^{-1}\rm\,cMpc$) of the \textsc{SIMBA} simulation.  We test this hypothesis by analysing the TNG-300-1 simulation in Appendix~\ref{app:Box&Res},  where we indeed find improved agreement with the \citet{Font-Ribera2013TheBOSS} observations for a larger box size of $300h^{-1}\rm\,cMpc$.

The main focus of this study are the differences caused by the various ionisation models.  In Fig.~\ref{fig:haloRadPhys} we find these differences are largest for the highest mass haloes with $M\geq 10^{12.8}\,M_{\odot}$. In general, the \textit{NoCol} and \textit{LoSo} models show less and more \lya\ transmission relative to the fiducial TNG100-1 model, respectively.   In the \textit{NoCol} model (orange curves) this is due to neglecting collisional ionisation from hot circumgalactic gas, where in general, the temperature and physical extent of the hot gas increases with halo mass.   Interestingly, the \textit{NoCol} model predicts too little transmission in the highest mass bin relative to the \citet{Prochaska2013QuasarsQuasars} measurements, suggesting that collisional ionisation (and hence gas temperature) plays an important role in setting the \lya\ transmission at $b<1h^{-1}\rm\,cMpc$ \citep[see also][]{Sorini2018ADistance}.  

The increased transmission in the \textit{LoSo} model (green curves) due to enhanced ionisation by the proximity effect is also most pronounced in the $M\geq 10^{12.8}\,M_{\odot}$ bin, as these haloes are populated with the highest luminosity AGN in our model.\addlines*\footnote{If the AGN emission is preferentially beamed along the line of sight rather than in the transverse direction, our isotropic emission model will overestimate the impact of the proximity effect on the transmission profile.  Similarly, non-equilibrium photo-ionisation and light travel time effects due to flickering AGN emission may also result in gas that is less highly ionised \citep[e.g.][]{OppenheimerSchaye2013,Schmidt_2019}.}  Note, however, that in contrast to the \textit{NoCol} case (orange curves), the differences between the \textit{LoSo} (green curves) and the fiducial model (black curves) are largest at  $1h^{-1}\rm\,cMpc\leq b\leq 5h^{-1}\rm\,cMpc$.  This is because the enhanced photo-ionisation rate only begins to dominate over collisional ionisation at $b\gtrsim 1 h^{-1}\rm\,cMpc$.  By contrast, the haloes in the lower two mass bins host either fainter AGN, or are unoccupied.  As a result, the local source model does not have a significant effect on the  \lya\ transmission profiles for haloes with masses $M \lesssim 10^{12.8}\rm\,M_{\odot}$.  Note, however, that we have deliberately adopted a model that maximises the proximity effect around the most massive haloes, and adopting a duty cycle $f_{\rm duty}<1$ \citep[e.g.][]{Shankar_2010} would push these AGN into lower mass hosts.  Finally, in the $M\geq 10^{12.8}\rm\,M_{\odot}$ bin the \textit{LoSo} model is in slightly better agreement with the \citet{Mukae20203DTechnique} data at $b<5h^{-1}\rm\,cMpc$, although due to the large error bars the significance is not high.  This appears to be consistent with the interpretation advanced by \citet{Mukae20203DTechnique} that the MAMMOTH1-QSO tomographic map exhibits a QSO proxmity zone. 

Since the \textit{LoSo} and \textit{NoCol} models effectively bracket the plausible range in the \lya\ transmission profiles, we proceed to investigate the effect these models have on the \lya\ transmission associated with protoclusters in TNG100-1.  The different sub-grid physics implementation in Illustris-1 sits between the extremes explored by these models, and so we do not investigate it further.


\section{The effect of local ionisation on the \texorpdfstring{L\lowercase{y}-$\alpha$}{} transmission around protoclusters}\label{sec:FvsGal}
\subsection{Smoothed \texorpdfstring{L\lowercase{y}-$\alpha$}{} forest transmission maps} \label{sec:ideal_maps}

\begin{figure*}
	\includegraphics[width=\textwidth]{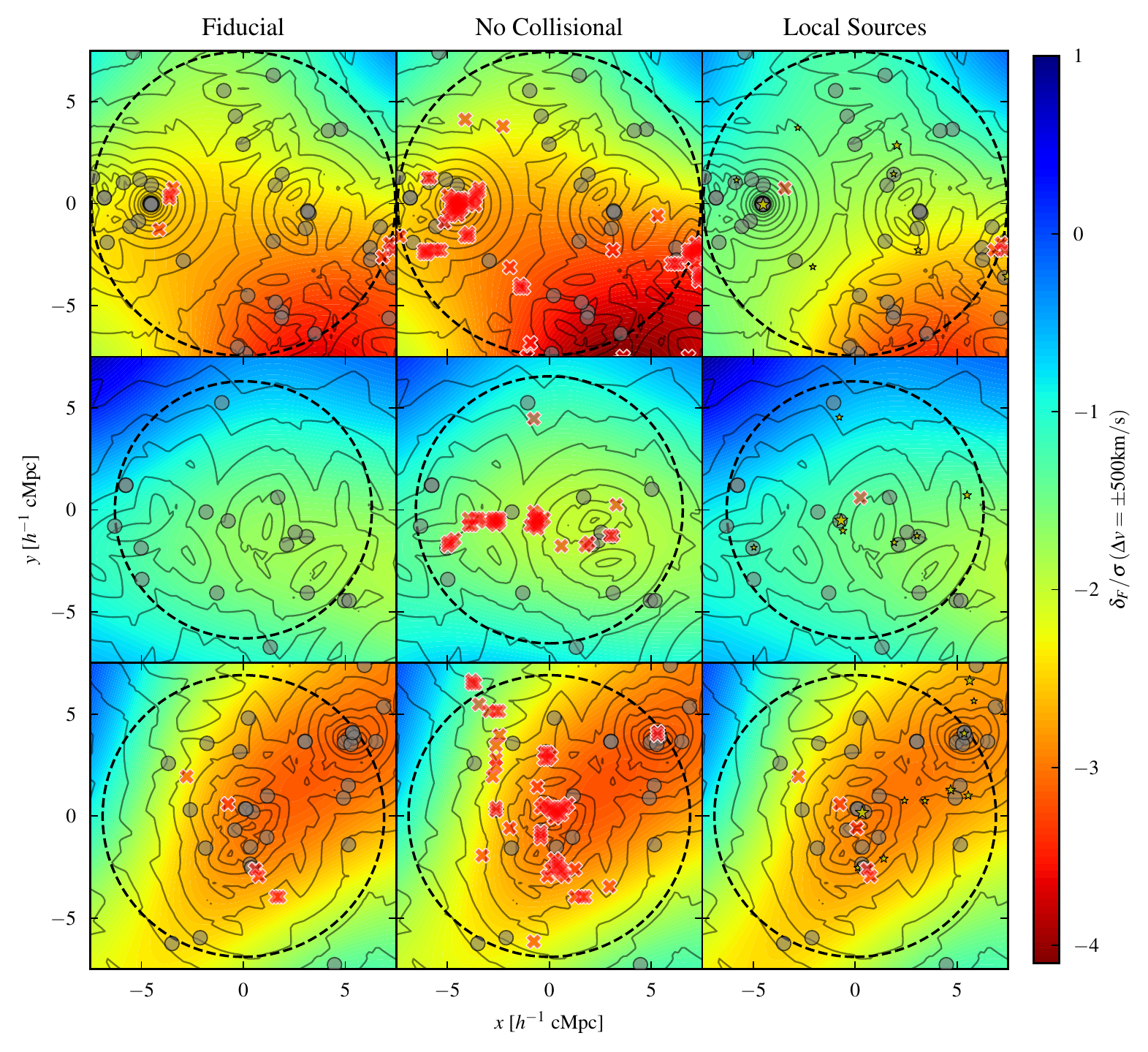}
	\vspace{-0.8cm}
    \caption{Smoothed transmission maps showing $\delta_{\rm F}/\sigma$ for three protoclusters in the TNG100-1 simulation at redshift $z=2.44$.  Positive (negative) values of $\delta_{\rm F}/\sigma$ correspond to \lya\ transmission that is larger (smaller) relative to the average for the IGM.  The maps are obtained by averaging the \lya\ forest over a $\Delta v= 1000\rm\,km\,s^{-1}$ window, smoothing in the transverse direction with a Gaussian filter with standard deviation $4h^{-1}\rm\,cMpc$, and then centring on velocity windows $\Delta v = \pm 500\rm\,km\,s^{-1}$  where $\delta_{\rm F}/\sigma$ is minimised within $R_{95}$ for each protocluster.  Each row shows a different protocluster for the fiducial (left), no collisional (centre) and local sources (right) ionisation models.  On top of each map we display the locations of coeval \lya\ emitting galaxies that satisfy the criteria $L_{\rm Ly\alpha}>10^{41.5} \rm\, erg\,s^{-1}$ and $\rm EW_{\rm Ly\alpha}>15\,$\AA\ (grey filled circles) using a simple empirical model (see text for details).  The grey contours show the logarithm of the LAE overdensity, $\log(1+\delta_{\rm LAE})=\log(\rho_{\rm LAE}/\langle \rho_{\rm LAE}\rangle)$, obtained from the distance to the 5th nearest neighbour in increments of $0.2\rm\,dex$.  In the right column, the yellow stars show the locations of AGN in the local sources model, with sizes scaled according to their luminosity. Red crosses display the locations of coherently strong Ly$\alpha$ absorption systems (CoSLAs).  The dashed black circle shows
    the 2D cross-section of the sphere of radius $R_{95}$ -- the radius that contains 95 per cent of the $z=0$ mass, $M_{\rm z=0}$ -- that intersects the velocity window for each protocluster. From top to bottom, the selected protoclusters have $M_{\rm z=0}=10^{14.18}\,M_\odot$, $M_{\rm z=0}=10^{14.46}\,M_\odot$ and $M_{\rm z=0}=10^{14.43}\,M_\odot$.  Note the protocluster in the middle row is also shown in Fig.~\ref{fig:PC_Map}.}
    \label{fig:pcMaps600}
\end{figure*}

We now turn to investigate how the \lya\ transmission around protoclusters is altered by changes in the local ionisation state of the IGM.    As already discussed , we do not create the \lya\ transmission maps by forward modelling the observational data \citep[e.g.][]{Stark2015ProtoclusterMaps}.   Instead, we use the noiseless spectra drawn from the simulations to create idealised maps of the relative transmission, $\delta_{\rm F}$, around each of the $22$ protoclusters in the TNG100-1 volume.  We then degrade these maps to match the final resolution of the observational data presented by \citet{Lee2018First2.55} and \citet{Newman2020LATIS:Survey} by smoothing with a Gaussian filter.  Our results will therefore not capture the effect of any systematic uncertainties associated with the accuracy of tomographic reconstruction techniques, or the signal-to-noise properties of the data. 

We first extract spectra in a $90\times 90$ grid in a $15~h^{-2}~\rm cMpc^2$ area centred on each protocluster's centre of mass, following the procedure described in Section~\ref{sec:spec}.  We then construct \lya\ transmission maps by obtaining the average \lya\ transmission over velocity windows, $\Delta v = 1000\rm\,km\,s^{-1}$ and then smoothing the relative transmission, $\delta_{\rm F}$, in the transverse direction using a Gaussian with standard deviation $4~h^{-1}\rm\,cMpc$.   The velocity window, $\Delta v$, is chosen to match the full width at half maximum of the Gaussian filter at $z=2.44$.  This choice matches the transverse smoothing scale applied in the \lya\ Tomography IMACS Survey \citep[LATIS,][]{Newman2020LATIS:Survey} and COSMOS \lya\ Mapping and Observations survey \citep[CLAMATO,][]{Lee2018First2.55} tomographic surveys.  Finally, we normalise each transmission map by the standard deviation of $\delta_{\rm F}$ obtained from the full simulation volume.  We obtain a standard deviation of $\sigma=0.076$, $\sigma=0.081$ and $\sigma=0.070$ for the fiducial, \textit{NoCol} and \textit{LoSo} models, respectively. The standard deviation is slightly increased in the \textit{NoCol} model with respect to fiducial, because ignoring collisional ionisation decreases the neutral hydrogen fraction in dense, hot gas, thus increasing the amount of strong \lya\ absorption.  Conversely, the standard deviation is reduced relative to fiducial in our \textit{LoSo} model, as the ionising sources (AGN) are placed into high density regions, thus reducing the incidence of strong \lya\ absorption.

The resulting \lya\ transmission maps for three different protoclusters are shown in each row of Fig.~\ref{fig:pcMaps600}.  The different local ionisation models for the protoclusters are displayed in each column, with the yellow stars in the right column showing the location of coeval AGN in the local sources model.   The three protoclusters have been selected to show: the region containing the most massive halo (and hence also the brightest AGN) in the TNG100-1 simulation (upper row, $M_{\rm z=0}=10^{14.18}\,M_{\odot}$), a region  where the \lya\ transmission within $R_{95}$ is higher than average (middle row, $M_{\rm z=0}=10^{14.46}\,M_{\odot}$) and a region that is  representative of the average \lya\ transmission associated with a protocluster in TNG100-1 (lower row, $M_{\rm z=0}=10^{14.43}\,M_{\odot}$).  Note the protocluster in the middle row is also displayed in Fig.~\ref{fig:PC_Map}.
The position of the maps are selected using the velocity window within $R_{95}$ where $\delta_{\rm F}/\sigma$ is minimised, similar to how these structures are identified within observed tomographic maps.     

In each map we also mark the locations of individual sight lines that contain coherently strong \lya\ absorption systems (CoSLAs) using red crosses.  Following \citet{Cai2017MAppingZ=2.32}, CoSLAs are defined as sight lines that exhibit a fluctuation in the \lya\ forest effective optical depth, $\delta_{\tau_\textrm{eff}} > 3.5$, over a scale of $15~h^{-1}~\rm cMpc$, after excluding any \lya\ absorbers with damping wings,  $\rm N_{\rm HI} \geq 10^{19}~\rm cm^{-2}$ (see also Paper I for further details).  This allows us to assess how CoSLAs are distributed relative to the \lya\ transmission maps. 

Lastly, we also use a simple model based on empirically derived scaling relations to display the locations of coeval Ly$\alpha$ emitting galaxies (grey circles).  The Ly$\alpha$ luminosities and equivalent widths for the galaxies were estimated using the stellar mass and star formation rate (SFR) for each sub-halo in TNG100-1.  We convert the instantaneous SFR into an luminosity at $1216$\,\AA\  using the relation from \citet{Dijkstra2017LectureWinterschool} for a \citet{Salpeter1955} initial mass function,
\begin{equation}
    L_{\rm Ly\alpha} = 1.0\times10^{42}\, \rm erg\ s^{-1} f^{\rm Ly\alpha}_\mathrm{esc} \left(\frac{\rm SFR}{M_\odot \,\rm yr^{-1}}\right). 
    \label{eq:Llya}
\end{equation} 
We have implicitly assumed a Lyman continuum escape fraction $f_{\rm esc}^{\rm LyC}\simeq 0$ in Eq.~(\ref{eq:Llya}), and $f^{\rm Ly\alpha}_\textrm{esc}$ is the volume averaged effective Ly$\alpha$ escape fraction inferred by \citet{Hayes2011OnGalaxies}, 
\begin{equation}
    f^{\rm Ly\alpha}_\mathrm{esc} = C_{\rm Ly\alpha}10^{-0.4\,A_{\rm Ly\alpha}},
    \label{eq:ColourExcess}
\end{equation}
where $C_{\rm Ly\alpha} = 0.445$ \citep{Hayes2011OnGalaxies}.  The quantity $A_{\rm Ly\alpha}$ is derived using the relation between extinction at H$\alpha$ wavelengths and stellar mass derived by \citet{Garn2010PredictingGalaxy}, and then converting to $A_{\rm Ly\alpha}$ using the \citet{Calzetti2000TheGalaxies} dust law.  We also estimate the rest frame equivalent width in Angstroms, $\rm EW_{\rm Ly\alpha}$, for each \lya\ emitter (LAE) using the relation $\rm EW_{Ly\alpha}=f_{\rm esc}^{\rm Ly\alpha}/0.0048$ from \citet{Sobral2019PredictingObservable}.  The LAEs displayed in the maps are selected by requiring $L_{\rm Ly\alpha}>10^{41.5} \rm\, erg\ s^{-1}$ and $\rm EW_{\rm Ly\alpha}>15\,$\AA\ \citep[e.g.][]{Shimakawa_2017}. The grey dashed contours correspond to the logarithm of the LAE overdensity, $\log(1+\delta_{\rm LAE})=\log(\rho_{\rm LAE}/\langle \rho_{\rm LAE}\rangle)$, determined by a fifth nearest neighbour algorithm.  We note, however, that this simple model does not include a self-consistent coupling between the visibility of the \lya\ emission line and the \lya\ opacity of the intervening circumgalactic medium (CGM) or IGM in the TNG100-1 simulation,  It furthermore does not follow the complex \lya\ radiative transfer within the interstellar medium of the galaxies \citep[e.g.][]{Laursen2011,GurungLopez2020}.  As such, while the model is consistent with average LAE properties by design, it may still underestimate the variation in $L_{\rm Ly\alpha}$ for a given stellar mass.

We first consider the fiducial ionisation model, displayed in the left column of Fig.~\ref{fig:pcMaps600}.  There is an anti-correlation between the LAE density and $\delta_{\rm F}/\sigma$ for all three protoclusters, and any CoSLAs are typically situated where the galaxy clustering is strongest.   The maps generally exhibit a smaller $\delta_{\rm F}/\sigma$ (less \lya\ transmission) where the LAE density is largest.  However, for the protocluster displayed in the top row of Fig.~\ref{fig:pcMaps600} there is an offset between where the LAEs are most strongly clustered around a massive halo with $M=10^{13.5}\,M_{\odot}$ at $(x,y)\simeq (-4,0)h^{-1}\rm\,cMpc$ and the largest \lya\ transmission decrement at $(x,y)\simeq (3,-6)h^{-1}\rm\,cMpc$.  This is qualitatively similar to the observation from \citet{Lee2016ShadowField}, where no strong \lya\ transmission decrement was detected around a galaxy overdensity in their CLAMATO tomographic maps.  These authors speculated that higher gas temperatures due to shocks or feedback may play a role in ionising gas and hence suppressing \lya\ absorption in the vicinity of galaxy overdensities.  This is indeed the case for the example here; the gas around the massive halo at $(x,y)\simeq (-4,0)h^{-1}\rm\,cMpc$ has been heated to $T>10^{6}\rm \,K$ and is therefore highly ionised, whereas the IGM associated with the \lya\ transmission decrement in the lower right of the map is significantly cooler, with $T <10^{5}\rm\, K$.  The protocluster displayed in the middle row (see also the same object in Fig.~\ref{fig:PC_Map}) exhibits more \lya\ transmission compared to the other protoclusters for a similar reason; in addition to the presence of larger underdensities within $R_{95}$ in this protocluster, there is an extended region of $T>10^{6}\rm\,K$ gas around the protocluster centre of mass that further increases the \lya\ transmission.

\begin{figure*}
    \centering
    \includegraphics[width=\textwidth]{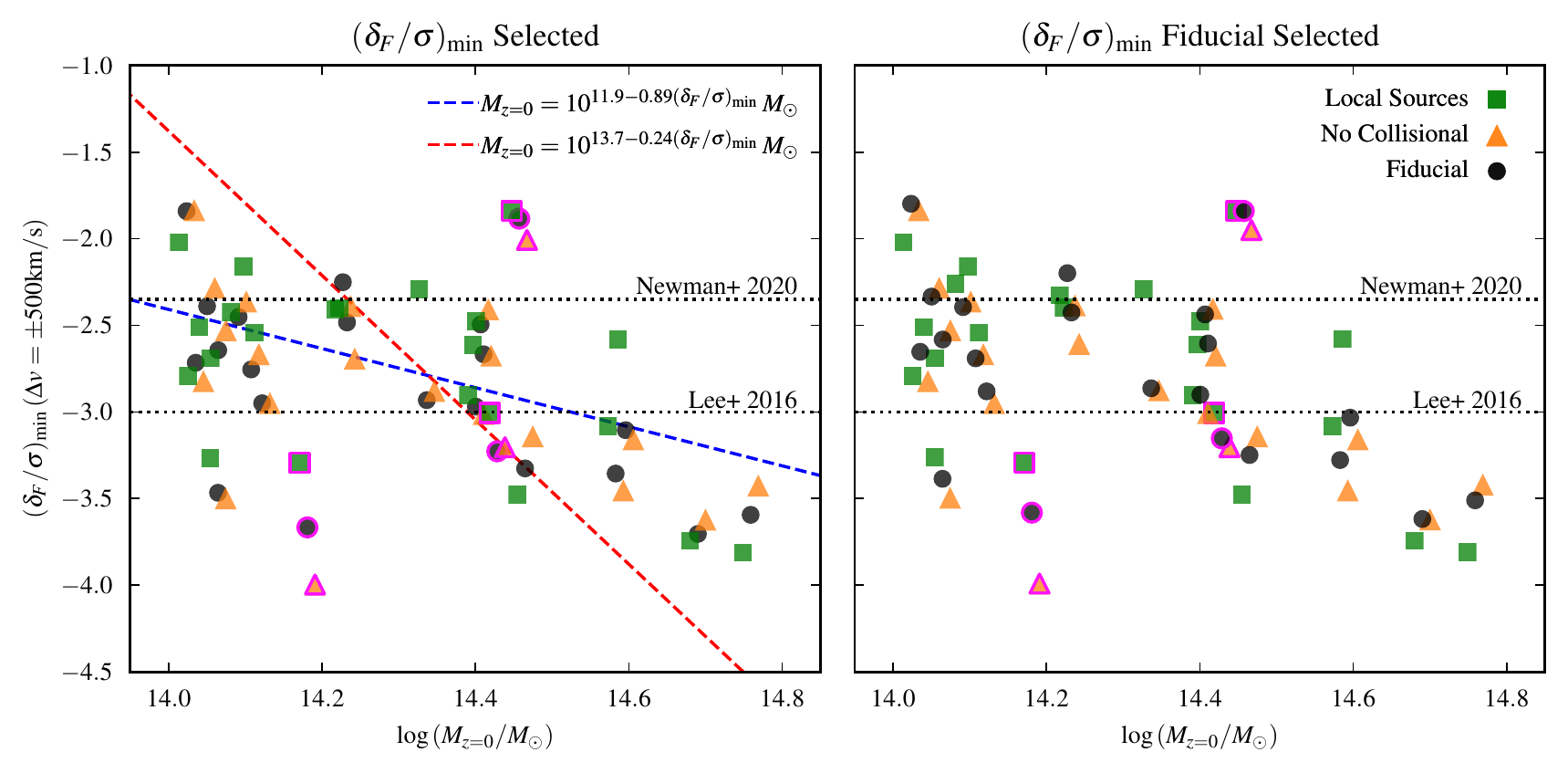}
    \vspace{-0.8cm}
    \caption{Scatter plot showing the relationship between the minimum relative \lya\ transmission, $(\delta_{\rm F}/\sigma)_{\rm min}$, and the $z=0$ cluster mass, $M_{\rm z=0}$, in the smoothed transmission maps for the 22 protoclusters in the TNG100-1 volume at redshift $z=2.44$.  We show the results for our fiducial (black circles), \textit{NoCol} (orange triangles) and \textit{LoSo} (green squares) ionisation models, where the \textit{NoCol} and \textit{LoSo} data points for each protocluster have been slightly offset on the horizontal axis for presentation purposes.  The dotted lines show the thresholds used by \citet{Newman2020LATIS:Survey} and \citet{Lee2016ShadowField} to select protoclusters from observed tomographic maps.  These are at $(\delta_{\rm F}/\sigma)_{\rm min} = -2.35$ and $-3$, respectively. The data points outlined in fuchsia correspond to the three protoclusters shown in Fig.~\ref{fig:pcMaps600}. The left panel shows the $(\delta_{\rm F}/\sigma)_{\rm min}$ identified within $R_{95}$ for each model.  The right panel instead shows $(\delta_{\rm F}/\sigma)_{\rm min}$ obtained when centring at the same location as $(\delta_{\rm F}/\sigma)_{\rm min}$ in the fiducial model, where we have also fixed $\sigma=\sigma_{\rm fid}$ for all the ionisation models. The red dashed line displays the relationship between $(\delta_{\rm F}/\sigma)_{\rm min}$ and $M_{\rm z=0}$ obtained by \citet{Lee2016ShadowField} from collisionless cosmological simulations post-processed with the fluctuating Gunn-Peterson approximation.   The blue dashed line shows a linear best fit to the data from our fiducial model.}
    \label{fig:minF-pcMass}
\end{figure*}

In the central column of Fig.~\ref{fig:pcMaps600} we show the smoothed \lya\ transmission maps for the same three protoclusters, but now using the \textit{NoCol} ionisation model.  As expected, all three protoclusters exhibit smaller values of $\delta_{\rm F}/\sigma$ where the LAE density is largest, but there is no significant change in  $\delta_{\rm F}/\sigma$ where the LAE density is lower.  This is because hot, collisionally ionised gas is found around massive haloes and filaments, and this is the environment where most of the LAEs reside in our model.  Another striking feature of the \textit{NoCol} ionisation models is that they contain a much higher incidence of CoSLAs (red crosses).  There are two reasons for this.  The first is that ignoring collisional ionisation produces larger \HI\ fractions, and hence stronger \lya\ absorption.  However, in the  \textit{NoCol} model we also neglect the effect of self-shielding to Lyman continuum photons on the \lya\ absorption.  This means that strong \lya\ absorbers with column densities $N_{\rm HI}>10^{19}\rm\,cm^{-2}$ in the fiducial model (i.e. damped systems) are over-ionised and have \emph{lower} column densities in the \textit{NoCol} model.  Hence, damped absorption systems that are excised when selecting the CoSLA sample in the fiducial model are erroneously classified as lower column density CoSLAs in the \textit{NoCol} model.   As discussed in Paper I, this highlights the importance of correctly modelling high column density absorbers when simulating the incidence of coherent \lya\ systems.

Finally, in the right column of Fig.~\ref{fig:pcMaps600} we show the smoothed \lya\ forest transmission maps for the \textit{LoSo} ionisation model.  The AGN positions are marked in the maps with star symbols.  Due to the proximity effect, all three protoclusters exhibit larger $\delta_{\rm F}/\sigma$ (more transmission) in comparison to the fiducial and \textit{NoCol} models.  The greatest increase in $\delta_{\rm F}/\sigma$ occurs where the LAE density is largest, but there is also a small increase in $\delta_{\rm F}/\sigma$ at lower densities.  This can be further understood from the halo profiles in Fig.\,\ref{fig:haloRadPhys}, where the brightest AGN in the model can ionise their surroundings up to $\sim5\,h^{-1}\,\rm cMpc$ from the centre of their host halo.  The largest proximity zone in the simulation volume is shown in the upper left panel of Fig.~\ref{fig:pcMaps600}, where the massive halo at $(x,y)\simeq (-4,0)h^{-1}\rm\,cMpc$ hosts an AGN with $M_{1450}=-24.7$.  This further enhances the existing spatial offset between the largest LAE density and the weakest \lya\ transmission/strongest \lya\ absorption.  A qualitatively similar observational result, but on much larger scales of $40h^{-1}\rm\,cMpc$, has been reported by \citet{Mukae20203DTechnique}, who find an \HI\ underdensity in the CLAMATO tomographic maps associated with a LAE overdensity.  These authors suggest this is due to the enhanced ionisation of the IGM by multiple nearby QSO proxmity regions.

Due to the increased level of ionisation around massive haloes, the \textit{LoSo} model also has a slightly reduced incidence of CoSLAs in comparision to the fiducial model.  Interestingly, however, there are a few cases in which a CoSLA is present in the \textit{LoSo} model but missing in the fiducial model.  An example of this can be seen in the central region of the protocluster in the middle row of Fig.~\ref{fig:pcMaps600}.  This is the result of absorption systems that are classified as damped ($N_{\rm HI}>10^{19}\rm\,cm^{-2}$) in the fiducial model and are thus rejected when selecting CoSLAs, but instead correspond to lower column density absorbers in the \textit{LoSo} model.  The column densities of the damped absorbers in the fiducial model are reduced due to the proximity effect, and these regions are then classified as CoSLAs.

\subsection{Protocluster masses and the correlation between LAEs and \texorpdfstring{L\lowercase{y}-$\alpha$}{} transmission in smoothed maps}

\begin{figure*}
	\includegraphics[width=\textwidth]{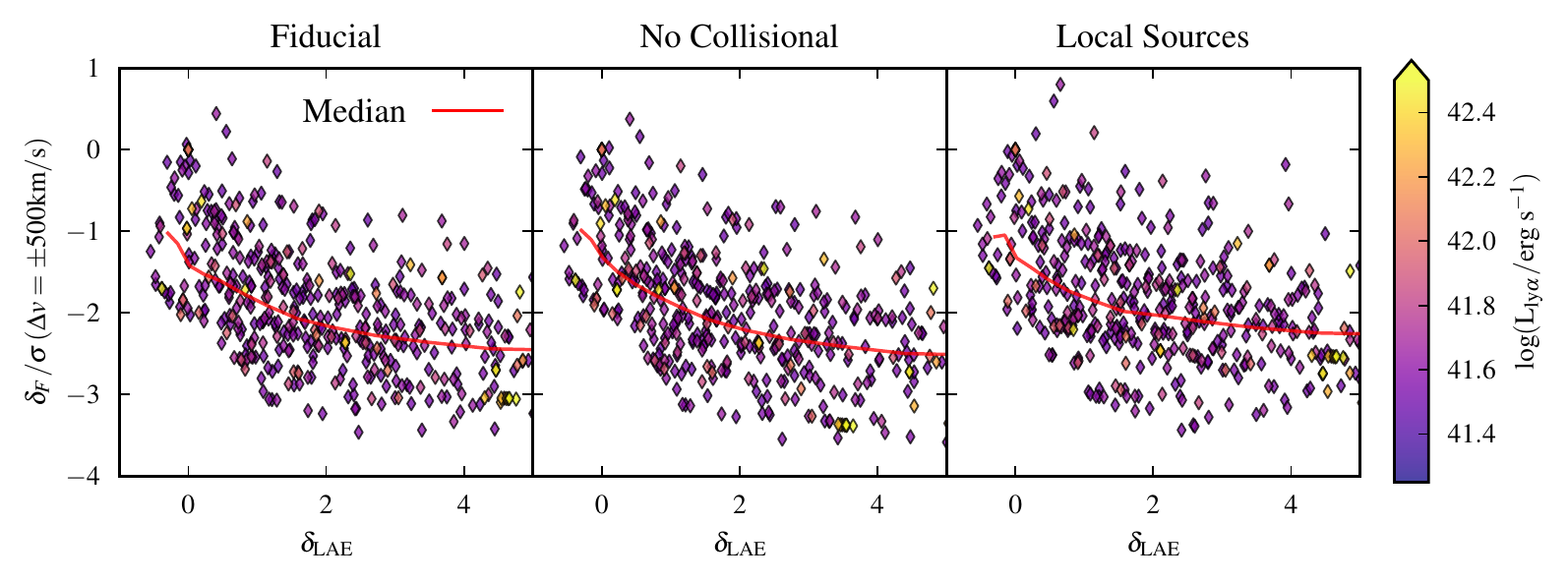}
		\vspace{-0.8cm}
	\caption{
	Scatter plot showing the relationship between the relative transmission, $\delta_{\rm F}/\sigma$, and the LAE overdensity, $\delta_{\rm LAE}$, in our smoothed transmission maps at redshift $z=2.44$.   From left to right, we show the results for our fiducial, no collisional and local source ionisation models.  The coloured diamonds correspond to LAEs with $L_{\rm Ly\alpha}>10^{41.5} \rm\, erg\ s^{-1}$ and $\rm EW_{\rm Ly\alpha}>15\,$\AA\ within $R_{95}$ of all $22$ protolusters in the TNG100-1 volume.   The colour scale shows the Ly-$\alpha$ luminosity of the LAEs, and the red curve shows the median trend obtained by randomly sampling within $R_{95}$ of the smoothed protocluster maps.  There is a weak anti-correlation between $\delta_{\rm F}$ and $\delta_{\rm LAE}$ for $\delta_{\rm LAE}\leq 1.5$. }
    \label{fig:LAE-vs-Gal_2000}
\end{figure*}

It is apparent from our qualitative discussion of the IGM transmission maps in Section~\ref{sec:ideal_maps} that local ionisation plays an important role in the correlation between \lya\ transmission, coeval galaxies and the distribution of coherent Lyman-alpha absorption systems within \emph{individual} protoclusters.  We now consider how local ionisation variations impact on two quantities derived from tomographic maps: estimates of the $z=0$ protocluster mass, $M_{\rm z=0}$ \citep{Lee2016ShadowField,Newman2020LATIS:Survey} and the correlation between LAE overdensity and \lya\ transmission \citep{Mukae2017CosmicField,Mukae20203DTechnique,Liang_2021}.

We first examine the relationship between the minimum relative transmission within the protocluster, $(\delta_{\rm F}/\sigma)_{\rm min}$, and the $z=0$ mass of the clusters, $M_{\rm z=0}$, in Fig.~\ref{fig:minF-pcMass}.  Candidate protoclusters are identified in both the CLAMATO and LATIS tomographic surveys by applying $(\delta_{\rm F}/\sigma)$ thresholds to the smoothed \lya\ transmission maps.  \citet{Newman2020LATIS:Survey} define their matter overdensity/protocluster candidates as regions with $\delta_{\rm F}/\sigma < {-2.35}$ in the LATIS survey, whilst for the CLAMATO survey \citet{Lee2016ShadowField} use a more conservative value of $\delta_{\rm F}/\sigma < {-3}$.

The left panel of Fig.~\ref{fig:minF-pcMass} shows $(\delta_{\rm F}/\sigma)_{\rm min}$ obtained from our smoothed transmission maps, centred on the velocity window containing $(\delta_{\rm F}/\sigma)_{\rm min}$ within $R_{95}$ for each protocluster. We find the wide variety of protocluster morphologies (see also Paper I) means the minimum transmission can be located anywhere up to $\sim 10h^{-1}\rm\,cMpc$ from the true protocluster centre of mass.  For comparison,  the right panel shows $(\delta_{\rm F}/\sigma)_{\rm min}$ when the maps for all models are instead centred at the location of $(\delta_{\rm F}/\sigma)_{\rm min}$ in the fiducial model. Additionally, in this case we use $\sigma = \sigma_{\rm fid}$ for all three models. This allows us to focus on how the ionisation models affect $\delta_{\rm F}$ in the same physical region, as opposed to selecting $\delta_{\rm F}/\sigma_{\rm min}$ in a way that mimics the observations.  
On average, the changes in $(\delta_{\rm F}/\sigma)_{\rm min}$ for the different local ionisation models are small in the left hand panel of Fig.~\ref{fig:minF-pcMass}, with the largest differences occurring for the protocluster with $M_{\rm z=0}=10^{14.22}\,M_{\odot}$ that harbours the brightest AGN/most massive halo in the TNG100-1 volume (see the upper panels of Fig.~\ref{fig:pcMaps600}).   This suggests that smoothing \lya\ tomographic maps on $\sim 4h^{-1}\rm\,cMpc$ scales should help mitigate for the possible bias in inferred cluster masses due to local ionisation variations, as well as optimising protocluster detectability  \citep{Stark2015ProtoclusterMaps}.  This is furthermore consistent with our earlier finding  that the largest differences in the local ionisation models occur on scales $<1h^{-1}\rm\,cMpc$ (see Fig.~\ref{fig:haloRadPhys}).   For comparison, we find that when selecting the same physical locations in the transmission maps and fixing $\sigma=\sigma_{\rm fid}$ (right hand panel), in almost all cases $(\delta_F/\sigma)_{\rm min}$ is largest in the \emph{LoSo} model and lowest in the \emph{NoCol}, as one would naively expect.  However, the differences between the ionisation models are again modest for most protoclusters.

From the left hand panel of Fig.~\ref{fig:minF-pcMass}, the protocluster completeness for the selection thresholds $(\delta_{\rm F}/\sigma)<2.35$ ($<3$) in the fiducial model is 86 (36) per cent, and this remains similar at 86 (41) per cent and 82 (32) per cent  for both the \textit{NoCol} and \textit{LoSo} models, respectively.   However, we find the best fit linear relation between $(\delta_{\rm F}/\sigma)_{\rm min}$ and $M_{\rm z=0}$ is slightly shallower compared to the relationship obtained by \citet{Lee2016ShadowField} from collisionless cosmological simulations (red dashed line in Fig.~\ref{fig:minF-pcMass}). Our best fit relation to the fiducial model is $M_{z=0}=10^{11.9-0.89(\delta_F/\sigma)_{\rm min}}\,M_\odot$, shown by the blue dashed line in the left panel of Fig.~\ref{fig:minF-pcMass}.  This implies that, for a given $(\delta_{\rm F}/\sigma)$, our fiducial model will favour larger $z=0$ masses for the most massive candidate protoclusters compared to the \citet{Lee2016ShadowField} calibration, possibly as a result of including hot gas with $T>10^{6}\rm\,K$ from shocks and AGN feedback (see also fig. 6 in \citet{Lee2016ShadowField} and the related discussion).  We caution, however, that the relatively small TNG100-1 box means we also have a much smaller sample of $M_{\rm z=0}>10^{14}\,M_{\odot}$ protoclusters compared to \citet{Lee2016ShadowField}, who use a collisionless dark matter simulation with box size $256h^{-1}\rm\,cMpc$.   Note also that in this work we analyse idealised transmission maps, and we have not performed a tomographic reconstruction of the \lya\ forest transmission using noisy data.

In the smoothed transmission maps in Fig.~\ref{fig:pcMaps600} we also observed an anti-correlation between the LAE overdensity, $\delta_\mathrm{LAE}$, and the relative transmission $\delta_{\rm F}/\sigma$.   In Fig.~\ref{fig:LAE-vs-Gal_2000} we examine this further by showing the relationship between $\delta_{\rm F}/\sigma$ and $\delta_{\rm LAE}$ for all 22 protoclusters in the TNG100-1 volume.  The three different ionisation models are shown in the individual panels.  The filled diamonds correspond to the LAEs in the transmission maps centred on $(\delta_{\rm F}/\sigma)_{\rm min}$, while the red curve shows the median relation obtained by randomly sampling the maps.  In all three models there is significant scatter in $\delta_F/\sigma$ at fixed $\delta_{\rm LAE}$, but the median trend shows decreasing $\delta_{\rm F}/\sigma$ with increasing $\delta_{\rm LAE}$ for $\delta_{\rm LAE}\lesssim 1.5$.  The Spearman's rank correlation coefficient for the LAEs with $\delta_{\rm LAE}<1.5$ is $-0.4$ in all three models, consistent with a weak anti-correlation.  This is followed by a flattening at $\delta_{\rm LAE}\gtrsim 1.5$ due to the $4h^{-1}\rm\,cMpc$ Gaussian smoothing we apply to the transmission maps; we have verified that adopting a smaller smoothing scale reduces this apparent flattening  and extends the anti-correlation to larger values of $\delta_{\rm LAE}$.   As was the case in Fig.~\ref{fig:minF-pcMass}, the relative transmission in the \textit{NoCol} and \textit{LoSo} models typically decreases and increases, respectively, compared to the fiducial model.  However, any changes remain very small compared to the scatter in the $\delta_{\rm F}/\sigma$--$\delta_{\rm LAE}$ plane, and are unimportant for the shape of the median trend.   The colours of each point in Fig.~\ref{fig:LAE-vs-Gal_2000} show the \lya\ luminosity of the LAEs, which are selected using the criteria $L_{\rm Ly\alpha}>10^{41.5} \rm\, erg\ s^{-1}$ and $\rm EW_{\rm Ly\alpha}>15\,$\AA.  There is no correlation (Spearman's rank coefficient $-0.04$ in all three models) apparent between $\delta_{\rm F}/\sigma$ and the LAE luminosity, $L_{\rm Ly\alpha}$ in our maps.

 We may also compare the results in Fig.~\ref{fig:LAE-vs-Gal_2000} to recent observational determinations of the relationship between $\delta_{\rm F}$ and $\delta_{\rm LAE}$ from \citet{Liang_2021} \citep[see also][for closely related work]{Mukae2017CosmicField,Mukae20203DTechnique,Momose_2020}, as well as the results from other cosmological hydrodynamical simulations \citep{Nagamine2020ProbingJWST}.  Note that these different studies do not calculate $\delta_{\rm LAE}$ in the same way as this work, so a direct comparison with the results we present here is not possible.  Nevertheless, we may still gain some insight from a qualitative comparison.  \citet{Liang_2021} identify LAEs at $z\sim 2.2$ from Subaru/Hyper Suprime-Cam data and compare the LAE overdensity to nearby \lya\ absorbers in the Extended-BOSS database \citep{Dawson_2016}.  These authors do not use a tomographic reconstruction of the \lya\ forest, and instead compute $\delta_{\rm LAE}$ and $\delta_{\rm F}$ within cylindrical apertures.  Assuming a best fit relation of $\delta_{\rm F}=m\delta_{\rm LAE} + C$, these authors find an anti-correlation with $m= -0.116^{+0.018}_{-0.022}$ and $C=-0.248^{+0.082}_{-0.093}$.  Similarly, \citet{Nagamine2020ProbingJWST} use the GADGET3-Osaka simulations to find a shallower relation with $m=-0.0664 \pm 0.00476$ and $C= -0.100 \pm 0.0006$, also obtained using a cylindrical aperture matched to the \citet{Liang_2021} measurement.  Although the slopes and normalisation of the linear fits from these two studies differ, the relationship between $\delta_{\rm F}$ and $\delta_{\rm LAE}$ is qualitatively similar to the weak anti-correlation we observe at $\delta_{\rm LAE}\lesssim 1.5$.  This is consistent with the interpretation that LAEs are preferentially located in regions with increased \lya\ absorption and hence larger \HI\ densities at $z\simeq 2$--$3$.  

Finally, although the relationship between $\delta_{\rm F}$ and $\delta_{\rm LAE}$ is not significantly altered in our different ionisation models, we note that the visibility of \lya\ emission lines and variations in the IGM/circumgalactic medium (CGM) \lya\ transmission are closely coupled.  As discussed previously, the volume averaged effective escape fraction we use, $f_{\rm esc}^{\rm Ly\alpha}$, does not self-consistently capture the effect of this coupling on $\delta_{\rm LAE}$ in the TNG100-1 simulation.  Detailed \lya\ radiative transfer models that include the effect of both inflows and outflows in the CGM will be required to investigate the relationship between $\delta_{\rm F}$ and $\delta_{\rm LAE}$ further \citep[e.g.][]{Barnes2011,Laursen2011,GurungLopez2020}


\section{Conclusions}
\label{sec:Conclusions}

In this work we have investigated the effect that local ionisation variations in the intergalactic medium (IGM), due the proximity effect from AGN and hot, $T>10^{6}\rm\,K$ gas from shocks and AGN feedback, have on the Ly-$\alpha$ absorption signature of protoclusters in the IllustrisTNG simulations at redshift $z\simeq 2.4$.    We consider three different local ionisation models in our analysis: a fiducial model with a spatially uniform UV background model, a second model that ignores the effect of collisional ionisation and self-shielding on the \HI\ fraction in the IGM, and final model where we incorporate spatial variations in the UV background due to the proxmity effect from AGN.   The impact of ``ionisation bias'' on the \lya\ transmission profiles around massive haloes and \lya\ transmission maps is then investigated.  We quantify this by computing the relative \lya\ transmission averaged over a velocity window $\Delta v$,  
$\delta_{\rm F}=(\langle F\rangle_{\Delta v}/\langle F \rangle) - 1$, where a negative (positive) value of $\delta_{\rm F}$ represents a decrease (increase) in the \lya\ transmission relative to the mean IGM transmitted flux, $\langle F \rangle$.  We furthermore examine the relationship between the relative \lya\ transmission in the smoothed transmission maps and the distribution of coeval \lya\ emitting galaxies (LAEs) and coherently strong \lya\ absorption systems (CoSLAs).  Our main conclusions are as follows:

\begin{itemize}
\item  We find local ionisation effects have a significant impact on \lya\ absorption in the vicinity of massive dark matter haloes with $M\geq 10^{12.8}\,M_{\odot}$ for impact parameters $b \lesssim 1\,h^{-1}\,\rm cMpc$ \citep[see also][]{Sorini2018ADistance}.  In particular, the presence of hot ($T>10^{6}\rm\,K)$ collisionally ionised gas will strongly increase $\delta_{\rm F}$ within $\sim1\,h^{-1}\,\rm cMpc$ of dark matter haloes. We furthermore find that the proximity effect associated with AGN (which have absolute magnitudes in the range $-24.7\leq M_{1450} \leq -18.9$ in our model) results in a modest increase in $\delta_{\rm F}$ for impact parameters, $1\,h^{-1}\,\textrm{cMpc}\leq b \leq5\,h^{-1}\,\textrm{cMpc}$, corresponding to distances where the photo-ionisation of the IGM begins to dominate over collisional ionisation.  However, both of these effects become less important on larger scales, $b\gtrsim5\,h^{-1}\,\rm cMpc$, and around less massive haloes with $M<10^{12.8}M_{\odot}$ in our model.\\

\item We construct idealised mock \lya\ transmission maps around the $22$ protoclusters with $M_{\rm z=0}\geq 10^{14}M_{\odot}$ in the TNG100-1 volume \citep[cf.][]{Lee2018First2.55,Newman2020LATIS:Survey}.  We find that local ionisation effects can play an important role in the correlation between the \lya\ transmission, coeval galaxies and CoSLAs within  a small number of individual protoclusters. 
In particular, we find a spatial offset of $\sim 9h^{-1}\rm\,cMpc$ between a LAE overdensity around a massive halo and the largest \lya\ flux decrement in a protocluster with $M_{\rm z=0}=10^{14.18}\,M_{\odot}$.   This offset is due to collisionally ionised gas with temperature $T>10^6\,\rm K$ surrounding the halo associated with the LAE density peak. This is qualitatively similar to the galaxy--\lya\ absorption offset observed by \citet{Lee2016ShadowField} and \citet{Mukae20203DTechnique} in CLAMATO tomographic maps.   The transmission contrast of this spatial offset is further enhanced by the proximity effect associated with a $M_{1450}=-24.7$ AGN hosted within the halo.   We furthermore find that the incidence of CoSLAs within protoclusters is sensitive to changes in our local ionisation models, largely as a result of changes in the number of self-shielded, damped \lya\ absorbers with $N_{\rm HI}\geq 10^{19}\rm\,cm^{-2}$ \citep[see also ][]{Miller2019SearchingAbsorption}.\\ 

\item After smoothing the simulated \lya\ transmission maps with a Gaussian of standard deviation $4\,h^{-1}\,\rm cMpc$ \citep{Lee2018First2.55,Newman2020LATIS:Survey} we find that local ionisation effects have a rather limited impact on the completeness of protocluster identification if using a fixed identification threshold of $(\delta_{\rm F}/\sigma)_{\rm min}\leq -2.35$ \citep{Newman2020LATIS:Survey} or $(\delta_{\rm F}/\sigma)_{\rm min}\leq -3.00$ \citep{Lee2016ShadowField}.  For an ensemble of $22$ protoclusters drawn from the TNG100-1 volume, we obtain a completeness of $82$--$86$ per cent and $32$--$41$ per cent, respectively for these thresholds if applied across all three of our ionisation models.  These results suggest that, in addition to optimising protocluster detection \citep{Stark2015ProtoclusterMaps},  smoothing the \lya\ tomographic maps on $4\,h^{-1}\,\rm cMpc$ scales may also help mitigate for a possible ``ionisation bias'' in the  completeness of a statistical sample of protoclusters.   Within our model, this is because the largest differences in the \lya\ forest transmission typically occur on scales $<4 h^{-1}\rm\,cMpc$ around dark matter haloes.   However, we also find the presence of hot gas around haloes may still result in systematically lower estimates of $M_{\rm z=0}$ for the most massive protoclusters if calibrating against mock tomographic \lya\ maps created using the fluctuating Gunn-Peterson approximation.  We find $M_{\rm z=0}=10^{11.9-0.89(\delta_{\rm F}/\sigma)_{\rm min}}\,M_{\odot}$ for the $22$ protoclusters in our fiducial model.\\

\item A simple model that uses empirically derived scaling relations for the volume averaged effective \lya\ escape fraction \citep{Hayes2011OnGalaxies} and the \lya\ rest frame equivalent width \citep{Sobral2019PredictingObservable} is used to populate the IGM transmission maps with \lya\ emitting galaxies.  In  qualitative  agreement with recent results from observations \citep{Mukae2017CosmicField,Liang_2021} and cosmological hydrodynamical simulations \citep{Nagamine2020ProbingJWST}, we observe a modest anti-correlation (Spearman's rank correlation coefficient of $\sim-0.4$) between $\delta_F$ and the LAE emitter overdensity, $\delta_{\rm LAE}$ for all three of our ionisation models at $\delta_{\rm LAE}\lesssim1.5$.  This is consistent with these galaxies being preferentially located in overdense regions which exhibit smaller $\delta_{\rm F}$ (i.e. stronger \lya\ absorption) at $z\simeq 2.4$ compared to the average IGM value.

\end{itemize}

\noindent
There remains plenty of scope for improving upon the numerical modelling in this work.  In particular, the dynamic range of the hydrodynamical simulations should ideally be larger.  A mass resolution of $M_{\rm gas} \sim 10^{6}\,M_{\odot}$ is required to resolve \lya\ absorption from the IGM at $z\simeq 2$ \citep{Bolton2009ResolvingSimulations,Miller2019SearchingAbsorption}.  While the \lya\ forest in the TNG100-1 simulation is therefore well resolved, the statistics are somewhat limited with only $22$ protoclusters with $M_{z=0}\geq10^{14}\,M_\odot$.  Additionally, the lack of any $M_{z=0}\geq10^{15}\,M_\odot$ clusters in the TNG100-1 simulation means that we are unable to study the impact of local ionisation effects on the most massive structures.   In our local sources model we have furthermore assumed isotropic AGN emission, and have ignored  non-equilibrium ionisation and light travel time effects \citep[e.g.][]{Schmidt_2019}.  Finally, we have adopted a simple model for the distribution of LAEs in the transmission maps that does not include the effect of the local IGM opacity on \lya\ emitter visibility.   Simulations of \lya\ radiative transfer through the interstellar and circumgalactic/intergalactic medium will be required to address this question further \citep[e.g.][]{Barnes2011,Laursen2011,GurungLopez2020}

In summary, models that incorporate all of these physical effects will be important for fully unravelling the relationship between \HI\ gas density and galaxies from \lya\ tomographic surveys at $z\gtrsim 2$.  Encouragingly, however, our results confirm that the  identification and completeness of $M_{\rm z=0}\simeq 10^{14}\,\rm M_{\odot}$ protoclusters identified from \lya\ forest transmission maps smoothed on scales $\gtrsim 4h^{-1}\rm\,cMpc$ should not be strongly affected by the variations in the local ionisation state of the IGM at $z\simeq 2.4$.  

\section*{Acknowledgements}

The IllustrisTNG simulations were undertaken with compute time awarded by the Gauss Centre for Supercomputing (GCS) under GCS Large-Scale Projects GCS-ILLU and GCS-DWAR on the GCS share of the supercomputer Hazel Hen at the High Performance Computing Center Stuttgart (HLRS), as well as on the machines of the Max Planck Computing and Data Facility (MPCDF) in Garching, Germany.  JSAM is supported by an STFC postgraduate studentship.  JSB acknowledges the support of a Royal Society University Research Fellowship. JSB and NH are also supported by STFC consolidated grant ST/T000171/1.

\section*{Data Availability}
All data and analysis code used in this work are available from the first author on reasonable request.  An open access preprint of the manuscript will be made available at arXiv.org


\bibliographystyle{mnras}
\bibliography{references.bib}


\appendix

\section{The effect of box size and mass resolution on \texorpdfstring{L\lowercase{y}-$\alpha$}{} absorption around haloes}

\label{app:Box&Res}

\begin{figure*}
	\includegraphics[width=\textwidth]{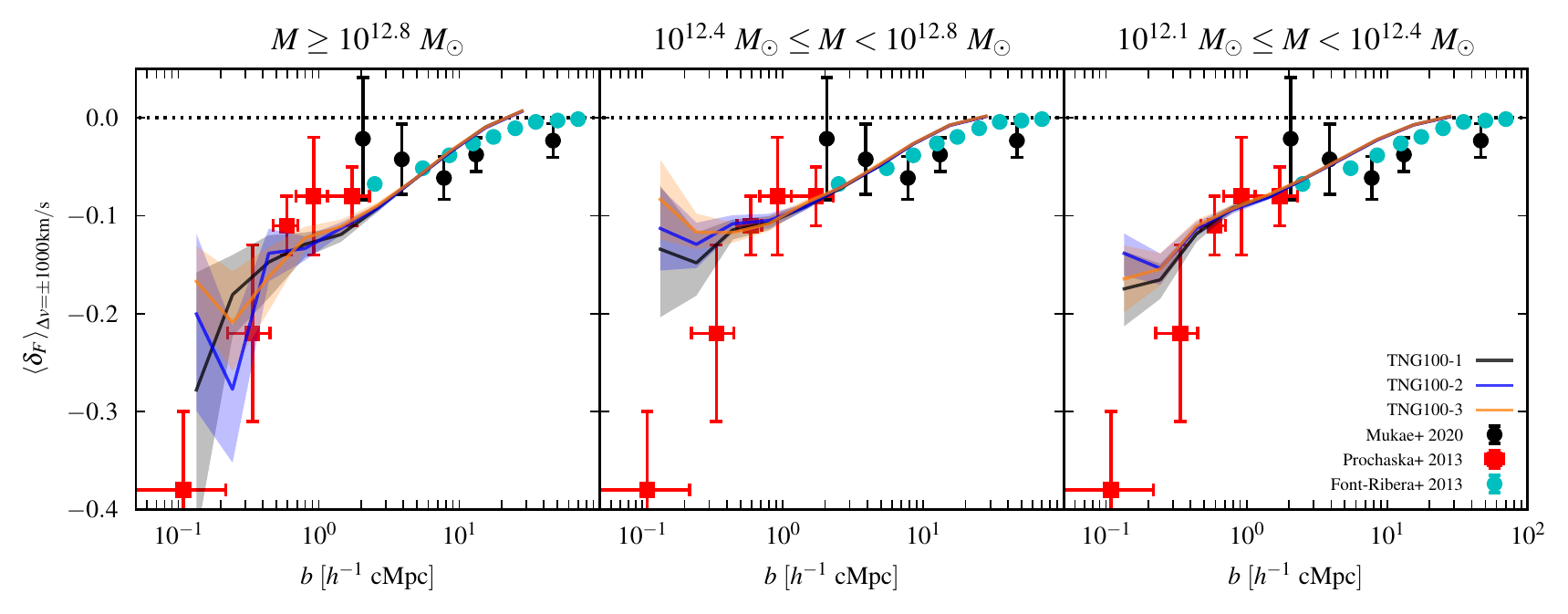}
		\vspace{-0.8cm}
    \caption{As for Fig.~\ref{fig:haloRadPhys}, but comparing simulations with fixed box size and different mass resolutions.   The fiducial TNG100-1 simulation (black curves) is compared to the TNG100-2 (blue curves) and TNG100-3 (orange curves) simulations.  These have dark matter particle masses a factor of $8$ and $64$ times larger than the TNG100-1 simulation, respectively.  The transmission profiles are consistent within the 68 per cent scatter around the median, shown by the shaded regions. }
    \label{fig:haloRadResComp}
\end{figure*}

\begin{figure*}
	\includegraphics[width=\textwidth]{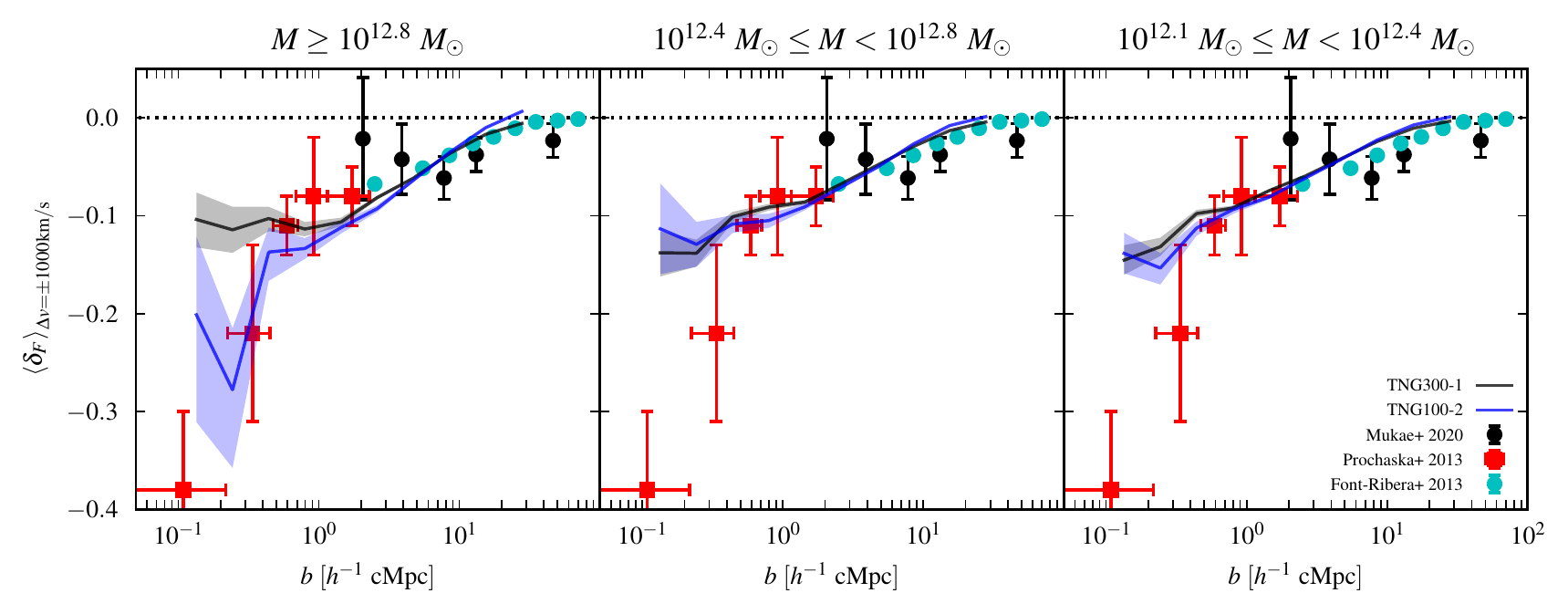}
		\vspace{-0.8cm}
    \caption{As for Fig.~\ref{fig:haloRadPhys}, but comparing simulations with different box sizes and very similar mass resolutions.   The TNG100-2 simulation (black curves) uses the same box size as the fiducial TNG100-1 model, whereas the TNG300-1 simulation (blue curves) has a volume $27$ times larger.  The TNG300-1 simulation is in slightly better agreement with the observational measurements from \citet{Font-Ribera2013TheBOSS} on large scales, particularly for the highest halo mass bin in the left panel.  The TNG300-1 model also exhibits more transmission at impact parameters $b<1h^{-1}\rm\,cMpc$ around the most massive haloes.}
    \label{fig:haloRadBoxComp}
\end{figure*}

Following on from Fig.~\ref{fig:haloRadPhys} and the associated discussion in Section~\ref{sec:haloes}, the effect of simulation mass resolution and box size on the transmission profiles around haloes are shown in Fig.~\ref{fig:haloRadResComp} and Fig.~\ref{fig:haloRadBoxComp}.  Here, in addition to the fidicial TNG100-1 model, we use the publicly available TNG100-2, TNG100-3 and TNG300-1 simulations. The properties of these additional simulations are outlined in Table~\ref{tab:Sims}.  There is generally very good agreement between the different simulations in Fig.~\ref{fig:haloRadResComp}, suggesting that our results should be sufficiently converged with respect to mass resolution.   However, for the simulations with varying box sizes in Fig.~\ref{fig:haloRadBoxComp} we observe larger differences.  In the case of the highest mass bin (left panel), there is a divergence between the two simulations at impact parameters $b<1\,h^{-1}\,\rm cMpc$.  This is caused by the larger number of massive haloes with $M\geq 10^{12.8}\rm\,M_{\odot}$ present in the TNG300-1 simulation, many of which are surrounded by hot $T>10^{6}\rm\,K$ gas with correspondingly low \HI\ fractions.  This explanation is consistent with the fact that in both of the lower mass bins we observe a very good agreement between the two simulations.  On larger scales ($b>1h^{-1}\rm\,cMpc$), the level of agreement  with the \citet{Font-Ribera2013TheBOSS} data is improved for the TNG300-1 model, particularly for the largest halo mass bin.  This appears to be consistent with the suggestion by \citet{Sorini2020Simba:Feedback} that smaller volumes lacking the most massive haloes may predict transmission that is systematically above the \citet{Font-Ribera2013TheBOSS} measurements.

\bsp	
\label{lastpage}
\end{document}